# Survey of Consensus Protocols

Abdul Wahab
*Faculty of Computer Science*
*Institute of Business Administration*
*Karachi, Pakistan*
abdulwahab@khi.iba.edu.pk

Waqas Memood
*Faculty of Computer Science*
*Institute of Business Administration*
*Karachi, Pakistan*
wmehmood@iba.edu.pk

*Abstract--* **Distributed ledger technology has gained wide popularity and adoption since the emergence of bitcoin in 2008 which is based on proof of work (PoW). It is a distributed, transparent and immutable database of records of all the transactions or events that have been shared and executed among the participants. All the transactions are verified and maintained by multiple nodes across a network without a central authority through a distributed cryptographic mechanism, a consensus protocol. It forms the core of this technology that not only validates the information appended to the ledger but also ensures the order in which it is appended across all the nodes. It is the foundation of its security, accountability and trust. While many researchers are working on improving the current protocol to be quantum resistant, fault-tolerant, and energy-efficient. Others are focused on developing different variants of the protocol, best suited for specific use cases. In this paper, we shall review different consensus protocols of distributed ledger technologies and their implementations. We shall also review their properties, concept and similar-work followed by a brief analysis.**

I. INTRODUCTION

Distributed ledger technologies have revolutionized the world by transforming the existing systems to become more secure, reliable and scalable. It forms a system that provides a trustworthy ledger among a group of nodes across a network that doesn't fully trust each other [1]. Distributed ledgers especially blockchain has been conceived as a provider of cryptocurrency but it has found its applications in different sectors including finance, academics, IoT, industries, and etc. That is why, we have witnessed an exponential adoption of this technology over the last few years. This has also raised the interest in the distributed ledger development community, which has scaled from hobbyists and academics to enterprises i.e. IBM and Intel. From the emergence of Bitcoin in 2008, there are currently many active development variants of this technology i.e. Ethereum, Hyperledger, Tangle, Corda, and etc. [2]

All these variants differ in the way they choose to reach the consensus, which helps a distributed ledger to function fairly, securely and efficiently. A consensus protocol, which is the core of the distributed ledger, performs two tasks: it guarantees that the next block of the network is the only version of the truth, and it protects the network from adversarial influences on the nodes and the network [1] [3]. It allows the network to confirm the transactions without relying on the intermediaries i.e. central authority. A consensus protocol makes a ledger functional and a flaw in the protocol will fail the accountability of the ledger. That is why, it owns a significant interest of the researchers and the industry. It also defines the nature of the distributed ledger which could be public, private or consortium/federated [4]. Another popular classification is permissioned and permissionless protocols.

**Public ledger** has no authority authorizing a transaction on the ledger. Anyone can join as a public node, validate transactions and participate in the consensus process without being permissioned. Transactions are public and transparent but the identity of the participants remain anonymous. **Private ledger** works with designated participants who are empowered to authorize transactions. Since data is unavailable for public view, it is ideal for implementation of data privacy rules and other regulatory compliance. However, this puts the system at the risk of security breaches just like in a centralized system for which it is argued that it is not a distributed ledger. Participants are identifiable in these systems but transactions remain encrypted and private. **Consortium ledger** is partially private ledger with the only difference lies in governance as the ledger is operated under the leadership of a group, not an entity. This way it provides all the benefits of the private chain without consolidating power to an individual and is also referred to as partial decentralized. [1] [4] [5]

The first consensus protocol of distributed ledger technology was proof-of-work (PoW) which powers the Bitcoin-Blockchain. It is based on a hash puzzle which is not only intensively resource consuming but also performs useless hashing. Also, it is not quantum proof and is subjected to various security threats of the future. That is why many different variants of the consensus protocols have been proposed and adopted. While some variants are the derivation of the existing protocols offering both minor and major adjustments, others propose an entirely different mechanism to reach the consensus. But collectively, they all are striving to develop an ideal fault-tolerant and resilient consensus protocol that not only protects the network from the present and future security attacks but also enforces an efficient and scalable implementation of its application.

The purpose of this paper is to provide an overview of some of the famous public, private and permissioned consensus protocols. The paper is organized as follows; To familiarize the readers with the subject, we will overview the history of the consensus protocol in Section II. In Section III, we will survey different consensus protocols, their properties, concept, implementations, analysis and use cases. We will also identify and discuss multiple variations of these protocols as similar work under the same section. Finally, in Section IV, we will conclude the paper by discussing how consensus protocols differ in their working and implementation which makes them ideal for specific use cases.

## II. BACKGROUND OF CONSENSUS PROTOCOL

Bitcoin is the first crypto-currency which was introduced in 2009 and uses proof-of-work as its consensus protocol [6]. The protocol forms the mining algorithm, avoids double spending and other attacks. The idea of a consensus protocol was inspired from the Adam Back's Hash cash, published as an improved revision of his previous paper in 2002, which proposed a protocol to prevent email spam and denial of service attacks [7]. In this section, we will visit the founding idea of proof-of-work, its evolution and different use cases till bitcoin.

The original idea of proof-of-work dates back to 1992, in which a group of authors presented a strategy to combat junk emails [8]. It utilizes different cost functions which are hard to compute and must be computed in order to gain access to the resources. This idea not only prevented the huge consumption of resources but also introduced the notion of using cryptography to increase the scarcity of a resource. The general idea of the paper was to bind each resource i.e. fax to a 'resource id', which can be obtained by computing a cost function. Unlike current proof of work model, it is not anonymous and saves user credentials for logging purpose.

Next idea came in the form of PayWord and MicroMint, two simple micropayments schemes, were presented by Ronald and Adi in 1996. They first discussed the faster computation of hash function as compared to public-key generation which is 10,000 times slower. Also, the verification of hash-based functions is 100 times faster as compared to the public-key cryptography [9]. The efficiency and performance of hash function over public-key cryptography techniques have welcomed possibilities for micropayments, which were not feasible earlier because the cryptography computation cost of the payment exceeds the payment itself. PayWord is a credit based micropayment scheme powered by a chain of hash values known as 'paywords'. MicroMint is another micropayment scheme implementation which is based on hash functions. The scheme comes with an exceptional performance but weaker security as compared to RSA based implementation.

In 1997, Matthew and Dahlia used the notion of proof-of-work for metering the popularity of the websites [10]. Website administrators can fraud the visit count of the website and charge higher rates for advertisement. By using robot programs or other tools, an individual can easily generate fake visit counts on a website. The paper introduced a timing function that is computed incrementally and whose result can be verified efficiently. Each visitor is asked to calculate a moderately hard cryptographic function to log its visit on the website server. To forge visit logs, a considerable amount of resource is required which is proportional to the amount of fraud. The output of the cryptographic function is stored on the server for accountability and auditability of website hits. The difficulty of the timing function which leads to resources consumption, security, accuracy and auditable metering mechanism.

Hashcash [11] implemented the proof-of-work protocol in terms of money that represents the burnt CPU cycle calculated as an n-bit partial hash collision. It was proposed as a software package by Adam Back in 1997 [11]. It offered a systematic countermeasure for denial of service attacks, junk emails and abuse of un-metered internet resources [11]. The protocol requires the client to compute a challenge to utilize the server resources. Partial hash collision requires the client to keep computing random hashes until a hash whose n-bits matches the target hash. It gives the control to arbitrarily alter the difficulty of the cost function by changing the number of bits to match, where one increasing bit makes the computing twice as difficult. As the machines get faster and efficient, the difficulty of the cost function can be adjusted accordingly. Its primary use case is to throttle the abuse of un-metered internet resources. To utilize a recourse or service i.e. email, the client must provide hashcash token, which results in an output of computing the cost function. An application of hash cash is to slow the spamming by generating a token for each email sent. Each token also has an expiry date to avoid accumulation of token and depleting resources all at once. Hashcash was further revised in 2002 [7].

Client Puzzle is another implementation of the proof-of-work protocol proposed in 1999, which utilizes cryptographic countermeasures to avoid denial of service attacks [12]. Two popular protocols, TCP and SSL, were under attack by leaving an unresolved large number of connections on the server. It would exhaust all the server resources thus making it incapable to serve legitimate requests. The attacks the server with TCP protocol targets the memory of the server but attacks on SSL protocol are more severe as it depletes the server's computational resources. Under Client Puzzle protocol, the server issues cryptographic puzzle to the new request. The new connections are only established when the client solves the cryptographic puzzle, which is hard to compute but easy to verify.

Hashcash [7] protocol requires the client to computes a token using the cost function to interact with the server. The cost function is termed as mint because of mining physical resources as a by-product. The aim of the challenge is to find n-bit collision against a fixed string. This n-bit matching, which was proposed earlier in 1997 [11], was further revised in this paper [7] to reduce the verification cost of the output. The protocol offers both interactive and non-interactive mode of communication. In interactive mode, the server generates a challenge for the client while in non-interactive mode, the client chooses his own random challenge and solves it. Each challenge has to be unique to avoid double spending of tokens. The server maintains a database of spent tokens to verify if it is an existing token and rejects requests with spent token. Also, each token that is generated as an output of computation, has an expiry date to avoid accumulation of tokens for later use. This avoids the chance of using all the accumulated tokens at once to launch a distributed denial of service attack in the future. The tokens are publicly auditable, as they can be efficiently verified by anyone.

All the above protocols enforce fair allocation of server resources among all users by avoiding massive server resource degradation during a DoS attack. The popular SYN-FLOOD attack was a major issue during the 1990s, as the attacker ends up consuming as many server resources as he can. Different characteristics of the discussed protocols and their evolution over time are the major inspiration behind the present-day consensus protocol of distributed ledger technology.

III. TYPES OF CONSENSUS PROTOCOL

A. Proof of Work

1) Introduction:

Proof of Work (PoW) [6] is the first consensus protocol for crypto-currency that allows the participant to reach consensus in the Bitcoin. The protocol is primarily based on costly computer computation involving Hashing (SHA-256), Merkle Tree and P2P networking for creating, broadcasting and verifying blocks on the network.

2) Properties:

- PoW is designed for **permissionless public** distributed ledger and consumes **computational resources** (or hashrate) of the system for mining.
- PoW maintains a block of transactions in a linear fashion (a single list of blocks) and each block contains a group of transactions.
- Each new block formation requires the miner to solve a cryptographic puzzle and the miner who solve it first, broadcasts its result to the network and takes the reward.
- This computational challenge-response process is called mining [6].
- Every transaction is cryptographically signed and the transaction is only accepted in the network only if the signature is valid and verifiable.
- In case of conflicts, the protocol extends multiple branches of blocks but only the longer branch is retained as the truthful branch.
- The protocol has a fair distribution of reward. The miner with p fraction of total computation power has a probability p to mine the next block.
- The management of the consensus is objective given that a new node can reach the current state of the network based solely on protocol rules. [13]

3) Concept:

PoW introduces the concept of mining which involves validation of a set of transactions (block) in the network by means of showing the computational proof of the work done. When a transaction is initiated, all the miners on the network race against each other to be the first to solve a cryptographic puzzle and create the block. The miner who successfully solves the puzzle, broadcasts his solution and block over the network to other peers, who after verification of the solution accept the new block on the chain. [6]

4) Implementations:

**Bitcoin.** Bitcoin [6] is the first and innovative peer to peer cryptocurrency that allows two parties to exchange payments without the need of an intermediator. Since its inception, it has not only revolutionized the financial industry but has also inspired other sectors i.e. health, management, governance and etc. Bitcoin provides payments exchange with trivial fees and identity anonymity. As a decentralized cryptocurrency, it is not influenced by the policies of the financial institution and avoids counterparty risk. Bitcoin allows micropayment channels through off-chain Lightening network [14] and native protocol library [15]. You can also sell computational data via zero-knowledge proof to attain utmost trust during a transaction [16]. It also supports multi-signature transaction over an address for improved security [17] [18].

**Litecoin**. Litecoin [19] is an open source and peer to peer cryptocurrency that is an implementation of proof of work [20]. Litecoin was forked from the Bitcoin codebase and went live in 2011. It uses improved security algorithms which is both computationally and memory intensive. It uses scyrpt [21] in its consensus protocol that makes it more expensive to counterfeit.

Other cryptocurrencies implementing PoW are Primecoin [22], ZCash, Monero, Vertcoin and etc.

*5) Analysis:*

- PoW is a power-hungry protocol that requires an immersive amount of computation power [23] which is a pure wastage of resources as we have new efficient protocols.
- The difficulty level in the PoW keeps increasing so as the power required to solve a harder cryptographic puzzle, which makes it inaccessible for solo miners to participate in the network.
- Due to its extensive power consumption, this protocol is considered a waste of huge resources. Other consensus protocols are recommended for efficient processing and better output [24].
- High computation requirement by the protocol also guarantees high security. A malicious user needs 51% of the computing power which is near impossible considering the computational difficulty level of the protocol. However, the protocol is highly vulnerable to Sybil and Denial of Service attack, and least affected by Selfish Mining attack. [25]
- The mining process of the system is also not fair; it is easily influenced by specialized hardware known as Application Specific Integrated Circuit (ASIC). These specialized and expensive machines give an unfair edge to miners over others in the network.

*6) Similar Work:*

**Proof of eXercise.** Proof of eXercise (PoX) [26] is designed for the public distributed ledger, consumes computational resources (or hashrate) of the system and is a conceptual consensus protocol that requires dedicated research for its practical implementation. It is an attempt to transform the PoW hash-based puzzle mining process towards forming a useful output and avoid wastages of resources. Currently, the annual electricity consumption of mining Bitcoin is equivalent to that of Ireland in 2014 [27], and is expected to be equal to that of the entire world by 2020 [23], thus making it unsustainable in the future [26] [24]. It proposes a variant of PoW that solves real-world scientific computation problems based on matrices as an eXercise. There are many real-world application of matrices based scientific problems including image processing, DNA and RNA matching and sequencing, data mining and etc.

**Proofs of Useful Work.** The nearly similar idea was published in 2017 as "Proofs of Useful Work", which proposes to solve the scientific problems based on orthogonal vectors OV as the useful proof-of-work [28]. The paper also integrated the idea of zero-knowledge proof [16] [28]. This enables the miners to give only the proof of the solution to their delegated task and not the solution itself to the delegator. The solution is made available by the network only when a particular pre-set condition is met. The paper does not discuss any challenges and practical implementation of the protocol.

## B. Proof of Stake

*1) Introduction:*

Proof of Stake (PoS) [29] [30] [31] is another consensus protocol that chooses the validator to mine the next block on the basis of its economic stake in the network (amount of coins a validator owns) and the age of that stake. PoS comes in many variants from minimal to significant changes in their base protocol. The most apparent fashion in which they differ is what strategy they implement to minimize the double spending and centralization issue in the protocol.

*2) Properties:*

- PoS is designed for **permissioned public** distributed ledger and works on **economically bonded** puzzle solutions.
- The process of computational challenge-response in the protocol is called minting.
- POS is weakly subjective given that a new node requires recent state, protocol messages, and rules to reach the current state of the network [13].
- As no new coins are generated, there is no block reward in the PoS and the miner only takes the transaction fee.
- The miner for a particular block is chosen in a deterministic way depending on its economic stake in the network.
- The protocol is also fair given that the probability p of a validator is directly proportional to the fraction p of the stake he owns out of all the in circulation.

*3) Concept:*

The PoS based ledger keeps track of all the validators (equivalent to miners in PoW) and their respective stake (cryptocurrency) in the network. In PoS, all the validators invest stake in the system to earn chances to mine the next block. The higher the stake, the higher the chances. However, it doesn't guarantee that the validator with the highest stake will be selected. The system chooses the validator randomly for block creation, like a lottery. If any participant tries to cheat the system, he loses his stake in the system. Unlike PoW, block creation is straightforward and doesn't require any significant computational power.

*4) Implementations:*

**Ethereum.** Ethereum [32] is an open-source and public blockchain powered by PoS as the protocol for reaching the consensus. Ethereum was initially PoW based cryptocurrency but it shifted its consensus mechanism to proof of stake, as it makes Ethereum energy efficient and secure as compared to proof of work counterparts [29]. It also features a powerful scripting language "smart contracts" to perform an operation on the blockchain. Ether is a cryptocurrency that works on the Ethereum blockchain. Unlike Bitcoin, which only offers peer to peer payment transfer, Ethereum offers a blockchain development stack on which developers can build and deploy DApps (Distributed Apps). This opens up the opportunity for developing unlimited ideas on this promising technology.

Other cryptocurrencies implementing PoS are Decred [33] [34], Peercoin [30], Neo, Navvcoin, Reddcoin, PivX, and Dash.

*5) Analysis:*

- PoS is an environment-friendly consensus protocol due to negligible computation requirement. Also, the protocol doesn't require any specialized hardware for participation.
- While PoS is energy efficient, it is more profitable for major stakeholders and biased towards.
- In PoS, an attacker would need 50%+ currency in the network to corrupt it, which is easier as compared to acquiring 51% computation power in its PoW counterpart. To avoid such security attacks, PoS design has several economic penalties to punish the colluding participant. This is indeed very effective since only major stakeholder can centralize the network and they will prohibit that to avoid penalty (losing the stake) by the network. The penalty is implemented in Ethereum whereas other implementations have attempted different strategies to solve this problem [13] [31].
- PoS has many security concerns, one of which is Bribe Attack [13]. This involves the process of reversing your own transaction for which an attacker builds his own chain and bribe the stakeholders for the block confirmations.
- PoS is attracting many new and existing implementations of distributed ledgers due to its energy efficient and decentralized design.

*6) Similar Work:*

**Delegated Proof of Stake.** Delegated Proof of Stake (DPoS) [35] is the most common variation of PoS, where stakeholders elect the validators rather than being the validator themselves. Unlike PoS, which follows the direct democracy, it works on the concept of representative democracy. Those who hold the wallet, vote for the validator to create the new block. Validators can also collaborate to create a block instead of competing against each other, unlike PoW and PoS. It boosts better distribution of reward as people tend to vote for the delegate (could be a casual user not necessarily rich) who will give back most rewards to them, thus favors decentralization. The voters make sure the honest behavior of the validator, whom they voted, to ensure the guarantee of the stake that they bet on the system. The downsides of the DPoS are 51% attack and cartels formation. Steem [36], EOS [37] and BitShares [38] are some of the popular implementations of DPoS.

**Leased Proof of Stake.** LPoS [39] is a less common yet enhanced variant of PoS focused on "richer gets richer" problem. It encourages the participants to lease out their stake to vote for the node. The node with the most stake has more chances to be allowed to create the new block. The reward received is then distributed among all the leasing participants as per the stake they bet. The system encourages number of leasing participants for the desire of rewards, thus improving the security of the protocol.

*7) Use Cases:*

- The technology is efficient and secure for the development of public crypto-currency. It is an ideal technology for the development of a public transaction system.

*C. Proof of Elapsed Time:*

*1) Introduction:*

Proof of Elapsed Time (PoET) [40] is another efficient consensus protocol that leverages the use of a Trusted Execution Environment (TEEs) i.e. Intel SGX-enabled CPUs [41] [42]. It extends both Proof of Time and Proof of Ownership to improve the efficiency of the mining process by following a fair lottery system. It leverages the capabilities of the TEEs platform to enforce random waiting time for block creation.

*2) Properties:*

- PoET is designed for **permissionless public** distributed ledger and utilizes Intel-based **specialized hardware** i.e. Intel SGX.
- The transactions and participants' time logs are transparent and verifiable which adds to the reliability of the network.

*3) Concept:*

The system is identical to PoW but it consumes far less computational resources. Unlike in PoW where the nodes compete against each other to solve a cryptographic puzzle and mine next block. In PoET, each validator is assigned a random wait time T for block construction which is assigned and monitored by the protocol. The first validator, whose finishes the waiting time (who has the shortest waiting time), creates and publishes his block on the network. The protocol works as the hybrid of first come first serve and random lottery fashion.

The code for the whole process relies on Software Guard Extension (SGX), available in most of the Intel CPUs, that ensures trusted code execution in a protected environment i.e. Intel Software Guard Extension. [42]

*4) Analysis:*

- Unlike other permissioned consensus protocol, PoET reaches consensus while maintaining the anonymity of the participants.
- The downside of the protocol is its dependency on TEEs enabled hardware. TEEs based hardware protect the system from malicious behaviors by maintaining a monotonic counter to ensure only one instance of the blockchain is running on the one CPU. This is important as participants may create multiple instances of wait time T to boost their luck. However, the protocol lacks security analysis and is vulnerable to different security attacks [43].
- Intel Software Guard Extension is susceptible to rollback [44] attacks and key extraction [45].

*5) Implementations:*

**Hyperledger Sawtooth.** Hyperledger Sawtooth [46] [1], developed by Intel, is a modular blockchain uses PoET consensus algorithm to implement a leader election lottery system. It offers parallel processing of the transactions for rapid block creation and validation by utilizing "Advance Transaction Execution Engine". It is an enterprise-grade protocol that is capable of huge throughput and large network population i.e. IoT network. The platform also provides development and execution of general purpose smart contracts on the ledger.

*6) Similar Work:*

**Proof of Luck.** Proof of Luck (PoL) is a permissioned and conceptual consensus protocol that is based on the use of a Trusted Execution Environment (TEEs) i.e. Intel SGX [41]. It extends both Proof of Time and Proof of Ownership to addresses the problem of extensive energy consumption [23] and centralization [47] of existing consensus protocol i.e. PoW and PoS. It also exhibits low transaction validation latency [28]. The block confirmation time for PoL is slightly greater than 15 seconds which is comparable to Ethereum (12 seconds on average) [48] and significantly less than of Bitcoin (10 minutes).

The protocol, for each round, signals participants to commit all the uncommitted transaction to a new block, and their version block is then given a numeric value. Then a voting is conducted where all the participants randomly vote for a number and the one with the highest votes wins (luckiest). The ledger accepts the luckiest block as the next block in the chain. Other participants stop mining and broadcasting their own block as soon as they receive the luckiest block to reduce network congestion.

*D. Proof of Space*

*1) Introduction:*

**Proof of Space (PoSpace)** [49], also known as proof of storage and **proof of capacity (PoC)** [50] [51], is an eco-friendly protocol that was initially formulated at [52] as an alternate idea to make a resource scared whose usage otherwise can be abused. PoSpace is a similar concept like PoW except it consumes disk storage instead of computation.

*2) Properties:*

- PoStorage/ PoC is designed for **public** distributed ledger and utilizes **free disk storage** as a resource.
- Influence of a miner's power over the network is directly proportional to the amount of disk space being contributed.

*3) Concept:*

**PoStorage/PoC** consumes disk space rather than computing resources to mine a block. Unlike PoW, where the miner keeps changing the block header and hash to find the solution, it generates all the random solutions, also called plots, using Shabal algorithm in advance and store it on the hard drive. This stage is called plotting and it may take days or even weeks depending on the storage capacity of your drive. Then on the next stage, miners match their solutions to the most recent puzzle and the node with the fastest solution gets to mine the next block. [53] [54]

*4) Implementations:*

**Burstcoin.** Burstcoin, an implementation of PoSpace/PoC, is a decentralized cryptocurrency and payment system that primarily relies on disk space as its mining resource. [50]. It was introduced by an anonymous developer in 2014 and is now being managed by the community. The mining in Burstcoin is inexpensive and efficient that it can even be performed on a mobile device. [55] It also led the first implementation of Turing Complete Smart Contract, which means that it can solve any reasonable computing problem. It is an active project and the community is currently working on its integration with the lightning network. This will not only decrease the transaction confirmation time but also increase the scalability of the blockchain system.

SpaceMint [56] is another conceptual cryptocurrency that is based on PoSpace.

5) *Analysis:*

- Mining can be performed on any ordinary hard drives and is not influenced by specialized hardware which makes it cheap. Additionally, miner doesn't need to upgrade their equipment and can utilize an older disk that can store data [53]. This makes it an ASIC proof [57] algorithm, which means that its performance can be significantly affected by hardware.
- Any user with free disk space even on their phones can participant in the network which leads to more decentralization.
- This technology could influence people to invest in larger disk spaces i.e. exabytes and zettabytes, that could ultimately lead us to another arm race [53].
- Hard drives consume approx. 30 times less power than an AISC based miner which makes the protocol energy-efficient.
- It requires more peer-to-peer interactivity than PoW which leads to network congestion.

E. *Proof of Retrievability*

1) *Introduction:*

Proof of Retreivability (PoR) is a consensus protocol that guarantees the existence of data on the peer by verifying the availability and integrity of small chunks of data [58]. Its application provides revolutionary solutions to Cloud Computing by enabling a peer to peer data storage network for data storage, transfer and retrieval.

2) *Properties:*

- PoR is designed for **public** distributed ledger and utilizes **free disk storage** as a resource.
- It is byzantine tolerant which makes it able to sustain adversaries and tolerate a class of failures known as Byzantine Generals' problem. [59]

3) *Concept:*

PoR provides standard functions to issue and verify the proof of files on a remote host i.e. peer. The protocol implements this via a challenge-response interaction called audit or heartbeat [60]. This helps the file owner (client) to verify the integrity and availability of the data over the host (peer). At first, the client wants to store a file F on the network N. The file is first encoded and then propagated over the network. The file is stored by multiple peers and the file owner/client can verify the integrity of the file by issuing challenges to the peers. The client receives the response to those challenges to verify the integrity. The encoding mechanism in the protocol plays an important part in the efficient processing of the data and verification of the chunks by the client.

4) *Implementations:*

**Storej.** Storej [60] is a peer to peer decentralized cloud storage network that guarantees data integrity. It allows participants and clients to share, store and transfer their data across different nodes. The data is protected from censorship, data loss and tampering by nodes for which they are rewarded. Storej leverages proof of and implements a challenge-response verification system that proves custody of a file for a participant. The nodes are required to maintain the client side security and encryption, and perform challenge puzzles to prove the integrity of data. Storej offers increases both data security and availability for enterprise data-centers which are very sensitive to data failures and breaches. This decentralized implementation is subjected to all the attacks that are common to almost all of the distributed system including some storage specific exploitations like Honest Geppetto.

5) *Analysis:*

- PoR is a compact proof by a client that it has the possession of a file. The compact proof incurs a low communication cost as only a random small chunk of the file is verified instead of transferring the whole file as a proof.
- The protocol also implements the encoding for processing the files on the network which are larger than the main memory of the peer (client).
- The protocol also gives multiple parameters to adjust trade-offs among security and performance.

F. *Practical Byzantine Fault Tolerant*

1) *Introduction:*

Practical Byzantine Fault Tolerant (PBFT) [59] extends and solves the classical problem of Byzantine Generals [61] in Distributed Computing. It states the fact that while reaching a consensus among multiple anonymous participants in a network, where all send their decisions to a leader and unanimously agree on that, there could be adversaries among those participants which pass the false message. It could potentially harm the network and leads to Byzantine failure. PBFT addresses these challenges while reaching a consensus in a 3-phase network intensive process.

*2) Properties:*

- PBFT is designed for **permissioned private** distributed ledger and consumes minimal **computational resources** (or hash rate) of the system.
- It can survive Byzantine faults in an asynchronous network by conducting block elections and maintaining leader nodes.
- The protocol can withstand malicious actors, hardware and software crash, and network failures.

*3) Concept:*

In PBFT, the objective of the protocol is to decide whether to accept a piece of information submitted to the ledger or not. Each node in the network maintains its own internal state and when it receives a message, it runs a computation and prepares a decision about the new message received. The individual decision of each node is sent to the leader of the nodes, which confirms the trust on the new message on the basis of the decisions from all the nodes. [59]

*4) Implementations:*

**Hyperledger Fabric.** Hyperledger [62] is a blockchain consortium under 'The Linux Foundation' and Fabric [63] [4] [1] is one of its popular implementations. It comprises a modular architecture and offers plug-and-play modules, where multiple implementations of the consensus protocol can be used by switching out the orderer module. The technology is backed by enterprise i.e. Intel, IBM, Cisco and etc., and offers private channels and smart contracts. [64] [65]

Ripple [66] and Stellar [67] are two other popular implementations of PBFT.

*5) Analysis:*

- The protocol can process high transaction throughput and can scale incredibly across the network.
- The protocol is highly efficient as it reaches the consensus asynchronously via the voting mechanism. This way of establishing consensus requires less effort than other methods but it comes at the cost of anonymity on the system.
- The protocol has centralization issues and can only be used in a permissioned fashion. A couple of selected nodes are allowed to form the consortium that makes it a semi-trusted network.

G. *Tangle*

*1) Introduction:*

Tangle [68] is considered as the next generation (a.k.a the third generation) of distributed ledger technology, an evolution to the blockchain. Tangle is a blockless blockchain based on Directed Acyclic Graph (DAG) [24], a finite directed graph with no directed cycles, which offers huge network scalability at low cost [69]. Hedera-Hashgraph [70] and Railblocks [71] are other famous implementations of DAG.

*2) Properties:*

- Tangle is designed for **permissionless public** distributed ledger and consumes minimal **computational resources** of the system for verification of the transactions.
- Unlike most of the protocols that use list type data structure, Tangle leverages the structural properties of the directed graph.
- It works in **blockless** manner and doesn't require to store the transactions in blocks nor it needs the order sequence of the previous transactions to keep the new in the right order. This means that all the transactions can be stored on multiple devices, and various locations.
- Transaction time is instantaneous; happens in a few seconds.

*3) Concept:*

For a transaction to get verified in Tangle, the issuer must verify two existing unapproved transactions (known as tips of the graph) on the ledger. This way transactions are verified simultaneously which offers scalability and faster transaction confirmation time. To avoid double spending problem, each incoming transaction randomly selects two unapproved transactions after running few simulations. These simulations provide us an estimate, confirmation confidence, of how many tips accepts this new transaction. Confirmation confidence for a transaction helps the protocol determine the validity of the transaction and its acceptability by the ledger.

*4) Implementations:*

**IOTA.** IOTA [68] [1], a cryptocurrency for Internet-of-Things (IoT) industry, was released in June 2016 and supports true nano-payments. It doesn't have any miner, block or transaction fee. It also introduces the concept of Data

Marketplace that permits participants to publish their data for paid subscription. It allows crowdsourcing of quality data for AI and machine learning.

*5) Analysis:*

- Tangle uses a SHA3 hash function variant, Kerl, utilizing ternary computations which make the protocol resistant to quantum attacks.
- High network scalability, transactional throughput and low transactional cost (computation) make are key highlights of the protocol.
- DAG based implementation, a ledger that stores transaction in an acyclic graph structure, adds efficiency and high throughput handling to the consensus protocol [24].
- Tangle is still working on offering a stable and mature implementation of smart contracts to ease the development of Decentralized Applications (DApps).
- It generates huge network congestion before it can start operating and any strategy to decrease that traffic renders network to vulnerable attacks.

*6) Use Cases:*

- Tangle, with no fee architecture, high scalability and throughput, is ideal for IoT projects that usually generate massive velocity and volume of data.
- Its architecture supports micro and low-cost transactions that makes it potentially favorite approach for future micropayment systems.
- Uber is a relevant example of a use case for tangle. Uber generates a huge volume of data and thousands of micro-transactions.

IV. CONCLUSION

Distributed ledger is a disruptive technology which has revolutionized the business processes with its application and adaptability. Behind every great distributed ledger implementation, there is a consensus protocol that powers it. In this paper, we surveyed a few popular consensus protocols. No consensus protocol is perfect, in fact, they all offer certain trade-offs among security, scalability efficiency and performance. Each of these protocols has its strengths and weaknesses, and all of them serve different purposes and provide domain specific solutions. But above all, they all exist to serve a common solution that is to prevent double spending in a distributed ledger.

The PoW introduced a decentralized payment system that offers security and data integrity at the cost of scalability and computational cost. It was then proposed to replace the useless work of PoW with a useful work i.e. solving a scientific problem instead of hashing. This remained a conceptual blockchain design and could witness a concrete implementation. Proof of stake (POS) effectively solved the problem of useless mining but the design introduced threats related to centralization. Its efficiency and secure implementation attracted many researchers and few PoW based implementations have also migrated to POS for example Ethereum. Presently, the trend is shifting towards a hybrid approach where an implementation is based on two or more consensus protocol e.g. Decreed [33] uses both PoW and POS.

The applicability and innovation of blockchain technology are not only limited to finance or academic sector. It is being continuously adapted and innovated to integrate with the bleeding-edge technologies. Tangle [68] is a chainless ledger technology that is built on the Directed Acyclic Graph (DAG). It can process thousands of transactions per second, offers a feeless architecture and can scale exponentially. Tangle is an excellent use case for IoT applications or any other system which requires to persist high velocity of data [24]. Likewise, Storej [60], based on PoR, solves the problem related to distributed data storage.

In conclusion, consensus protocol, the engine of a blockchain, comes in varied implementations and serves different use cases. Since the inception of PoW [6] in 2008, researchers around the world have been working to develop a secure, scalable and efficient consensus protocol which could produce tremendous results to the growth of economy and infrastructure.